\documentclass[reprint,superscriptaddress,prb]{revtex4-1}

\usepackage{amsmath}  
\usepackage{amsfonts} 
\usepackage{graphicx} 
\usepackage{hyperref}
\usepackage{amssymb}

\begin{document}

\title{The Pulsar Search Collaboratory: Current Status and Future Prospects}

\author{Harsha Blumer}
\email{harsha.blumer@mail.wvu.edu}
\author{Maura A. McLaughlin}
\author{John Stewart}
\author{Kathryn Williamson}
\author{Duncan~R.~Lorimer}
\affiliation{Department of Physics and Astronomy, West Virginia University, Morgantown, WV 26506, USA}
\altaffiliation{Center for Gravitational Waves \& Cosmology, West Virginia University, Chestnut Ridge Research Building, Morgantown, WV 26505, USA}
\author{Sue Ann Heatherly}
\affiliation{Green Bank Observatory, Green Bank, West Virginia, WV 24944, USA}
\author{Joseph K. Swiggum}
\affiliation{Center for Gravitation, Cosmology and Astrophysics, Department of Physics, University of Wisconsin-Milwaukee, P.O. Box 413, Milwaukee, WI 53201, USA}
\author{Ryan S. Lynch}
\affiliation{Green Bank Observatory, Green Bank, West Virginia, WV 24944, USA}
\author{Cabot Zabriskie}
\author{Natalia Lewandowska}
\affiliation{Department of Physics and Astronomy, West Virginia University, Morgantown, WV 26506, USA}
\author{Aubrey Roy}
\author{Shirley Au}
\affiliation{SmartStart Evaluation \& Research, 4482 Barranca Parkway, Suite 220, Irvine, CA 92604, USA}

\date{\today}

\begin{abstract}
The Pulsar Search Collaboratory (PSC) is a collaboration between the Green Bank Observatory and West Virginia University, funded by the National Science Foundation. The PSC program is currently expanding nationwide and engages high school students, teachers, and undergraduate mentors in real-world research by searching for
pulsars in data collected with the 100-m Green Bank Telescope. In the process, students learn about observational
radio astronomy, radio frequency interference, pulsar timing, and data analysis procedures. The primary goals of the PSC are to stimulate student interest in Science, Technology, Engineering, and Mathematics (STEM) careers, to prepare teachers in implementing authentic research with students by training them within a professional scientific community, and to promote student use of information technologies through online activities and workshops. In this paper, we provide an overview of pulsar science and the data analysis students undertake, as well as
a general overview of the program. We then discuss evaluation data collected from participants through a series of survey questions to determine if the program's initial goals were met. The program had a positive impact on the students according to multiple measures, in particular, on their understanding of the nature of scientific inquiry and motivation to pursue STEM career paths.

\end{abstract}

\maketitle

\section{Introduction} 

The National Science Foundation (2007) states, ``In the 21st century, scientific and technological innovations have become increasingly important as we face the benefits and challenges of both globalization and a knowledge-based economy. To succeed in this new information-based and highly technological society, all students need to develop their capabilities in Science, Technology, Engineering, and Mathematics (STEM) to levels much beyond what was considered acceptable in the past".  Students with advanced STEM degrees are required at a variety of skill and knowledge levels including information and medical technologies workers, high quality STEM teachers, and scientists and engineers working in research labs. An understanding of science is required to make informed decisions on political issues such as environmental regulations and to make sound personal consumer choices on health-care decisions. During the past century, STEM fields propelled United States to the forefront of an innovation-based global economy and, hence, there is a clear need to increase the number of students who obtain advanced degrees in STEM disciplines. To do this, it is essential to train K-12 teachers in STEM disciplines by supporting real research experiences and inquiry-based activities in classrooms. Recent studies show that students who had authentic scientific research experiences in high school, who undertook an apprenticed mentorship or internship, and whose teachers connected the content across different STEM courses were more likely to complete a STEM major than their peers who did not have these experiences (Roberts \& Wassersug 2009; Rosen et al. 2010).

The Pulsar Search Collaboratory (PSC) is one such comprehensive astronomy education program designed to build students' awareness and interest in STEM careers (Rosen et al. 2010). The PSC began as a joint project between the Green Bank Observatory (GBO) and West Virginia University (WVU), and is funded by the National Science
Foundation (NSF).  The program engages middle and high school students in analyzing real data collected using the Robert C. Byrd Green Bank Telescope (GBT) for the purpose of discovering pulsars. It also provides high school teachers with the experience of conducting scientific research.

The program helps students and teachers meet the Next Generation Science Standards (NGSS 2013) as follows:
\begin{enumerate}
\item Nature of Science/Science as Inquiry: The PSC provides students with experience in conducting scientific research, introducing core content and integrating the teaching of facts of science with the practice of scientific research.
 \item Content:
Pulsars (see Section \ref{sec:pulsar}) are fascinating objects for study, and research into their extreme properties has impacted many areas of fundamental physics. The PSC allows students to gain first-hand insights into several areas of the NGSS Physical Science Content Standards including motions and forces, conservation of energy, and interactions of energy and matter.
 \item
21st Century skills: The PSC requires students to work in teams, communicate their results to one another, and to use technology to improve their ability to do science.
\end{enumerate}

In this paper, we provide an overview of pulsar science (Section \ref{sec:pulsar}), origin of the PSC program (Section
\ref{sec:hist}), processing of radio telescope data (Section \ref{sec:data}), and goals and structure of the program (Section \ref{sec:psc}). Section \ref{sec:results} discusses results from the evaluation of the program.

\section{Why study pulsars?}
\label{sec:pulsar}

Pulsars are rapidly rotating, highly magnetized neutron stars born in the supernova explosions of massive stars. With masses $\sim$1--2 times that of our sun, diameter~$\sim$20 km, densities comparable to those inside atomic nuclei ($\sim$10$^{15}$ g~cm$^{-3}$), and rotation (spin) periods ranging from seconds to milliseconds, they are some of the most extreme objects in the universe. As a pulsar spins, charged particles are accelerated along the magnetic field lines causing the particles to radiate beams of radio waves which sweep across the sky. A pulse of radiation is seen each time the radio beam crosses the line of sight of an observer and can be detected by a radio telescope. The radio emission continues for $\sim$10--100 Myr until the pulsar eventually becomes a quiet neutron star.  A sub-class of pulsars, millisecond pulsars (MSPs), are short-period pulsars with periods of less than 30 ms. MSPs have been spun-up to high rotational frequencies by accumulating mass and angular momentum from a companion star. They have low magnetic fields and are expected to emit radio waves for billions of years. These objects are incredibly stable rotators with pulse arrival times measurable to microsecond precision and spin periods predictable to one part in 10$^{15}$ (McLaughlin 2013). Their very stable periods make them nearly ideal probes of a wide variety of astrophysical phenomena. These pulsars can be used to detect planets around pulsars, to test the accuracy of gravitational theories, and to establish pulsar-based timescales rivalling the accuracy of an atomic-clock (McLaughlin~2013).

The most transformative pulsar application attempts to directly detect low-frequency gravitational waves by regular monitoring a widely distributed array of MSPs that form a Galactic scale gravitational-wave observatory. Such a network of celestial clocks is called a Pulsar Timing Array (PTA). By precisely monitoring pulse arrival times, radio astronomers can determine the rotation period of a pulsar, how its rotation is slowing down, and whether the pulsar is orbiting a companion star. In addition, astronomers can measure how interstellar medium affects pulse propagation. If a gravitational wave passes through Earth, it will compress and stretch spacetime so that some pulsars' signals will arrive slightly earlier than normal and some will arrive slightly later. By looking for correlated changes in pulse arrival times across a network of many pulsars, astronomers in the North American Nanohertz Observatory for Gravitational Waves (NANOGrav; McLaughlin 2013) and International Pulsar Timing Array (IPTA -- a consortium of constituent PTA collaborations working together across continents; Manchester 2013), are poised to detect low-frequency gravitational waves from the most supermassive black holes at the centers of colliding galaxies billions of light years away. The most effective way to increase the sensitivity of this experiment is through the addition of pulsars to the PTA. PSC participants can contribute to this effort by discovering pulsars which may be added to the PTA.

\section{Origin of the PSC}
\label{sec:hist}

The 100-m GBT, with its fully steerable dish (the telescope can be pointed in any direction above the horizon), is one of the best telescopes in the world for finding and studying pulsars. During the summer of 2007, part of the GBT steering mechanism on which the 8 million kilogram structure rotates, was replaced.  While the repairs were being performed, the telescope was not fully steerable. The telescope normally could be rotated through two independent angles; while under repair it could only be rotated through one angle. This angle could only be changed occasionally
to allow the repairs to progress. Over this time, a team of astronomers carried out a survey in order to discover pulsars by locking the telescope at a constant angle with respect to the horizon and collecting data as the sky drifted through the telescope's aperture. Over the summer, this team surveyed over 25\% of the sky and amassed 130 TB (terabytes) of data at radio frequencies of 350 MHz. With a large amount of data to explore and discoveries to be made, this team of astronomers decided to involve teachers and high school students in the search for pulsars with a grant from the NSF (award \#0737641). The data collected during this period, which would form the data in the PSC database, would only be examined by students, not trained astronomers. As such, any discoveries made would belong to the PSC students. The PSC program thus began in 2008 with the goals of advancing science teachers' and students' understanding of the nature of science, preparing teachers to implement authentic research with their students, and
increasing student interest in STEM fields by apprenticing them within a professional scientific community analyzing data to search for pulsars.

\section{Data processing}
\label{sec:data}

The data collected by the telescope must be processed before they can be analyzed. We use PRESTO (\url{https://www.cv.nrao.edu/~sransom/presto/}) to process the data for PSC participants, the same processing software that produces pulsar diagnostic plots for professional astronomers. These plots are stored in the PSC database (\url{http://psrsearch.wvu.edu}) and become available to PSC participants after training. 

The GBT collects continuous radio frequency data in 2048 frequency channels. As the Earth turns, the sky passes over the aperture of the telescope. It takes about two minutes for a star to pass through the aperture. The data are first divided into two-minute segments, called ``pointings". Each pointing is processed independently. First, the man-made sources of radio waves or radio frequency interferences (RFIs) are removed.  Next, the data are corrected for the effects of dispersion of interstellar medium (ISM) and converted into a time series. The free electrons in the ISM disperse signals from a pulsar, causing the lower frequency ($\nu_{lo}$) signals to arrive at Earth after the higher frequency ($\nu_{hi}$ ) signals. This manifests itself as a ``smearing" of an otherwise sharp pulse when a pulsar is observed over a finite band width. Therefore, the effects of dispersion smearing should be removed through a process called ``de-dispersion". This involves applying a time shift to the signal at some given frequency. The amount of time shift in arrival times of a signal (or dispersive delay, $\Delta$t) between two observing frequencies (measured in GHz) is given by Eqn.~\ref{eqn:delay},
\begin{equation}
\label{eqn:delay}
\small
\Delta t=4.15~\mbox{ms} \times \left[ \left( \frac{\nu^{-2}_{lo}}{GHz}\right) -\left( \frac{\nu^{-2}_{hi}}{GHz}\right)\right] \times \left(\frac{\mbox{DM}}{cm^{-3}pc}\right)
\end{equation}
and is proportional to the dispersion measure (DM). DM is the integrated line-of-sight density of free electrons between an observer and the pulsar, given by Eqn.~\ref{eqn:dm},
\begin{equation}
\label{eqn:dm}
\mbox{DM}=\int^d_0 n_e dl
\end{equation}
where $n_e$ is the density of electrons and $d$ is the distance to the pulsar. DM is proportional to the distance that the signal has traveled and is expressed in units of parsec per centimeter-cubed (pc~cm$^{-3}$). Since DM of a source is unknown a priori, it is necessary to de-disperse the signal at many trial values which is done by appropriately shifting the frequency channels in time and summing them to form a ``de-dispersed time series". If this frequency-dependent delay is not accounted for, the summed pulse would appear blurred and smeared, or may even become undetectable.
 
\begin{figure*}[th]
 \label{fig:plot}
\includegraphics[width=\textwidth]{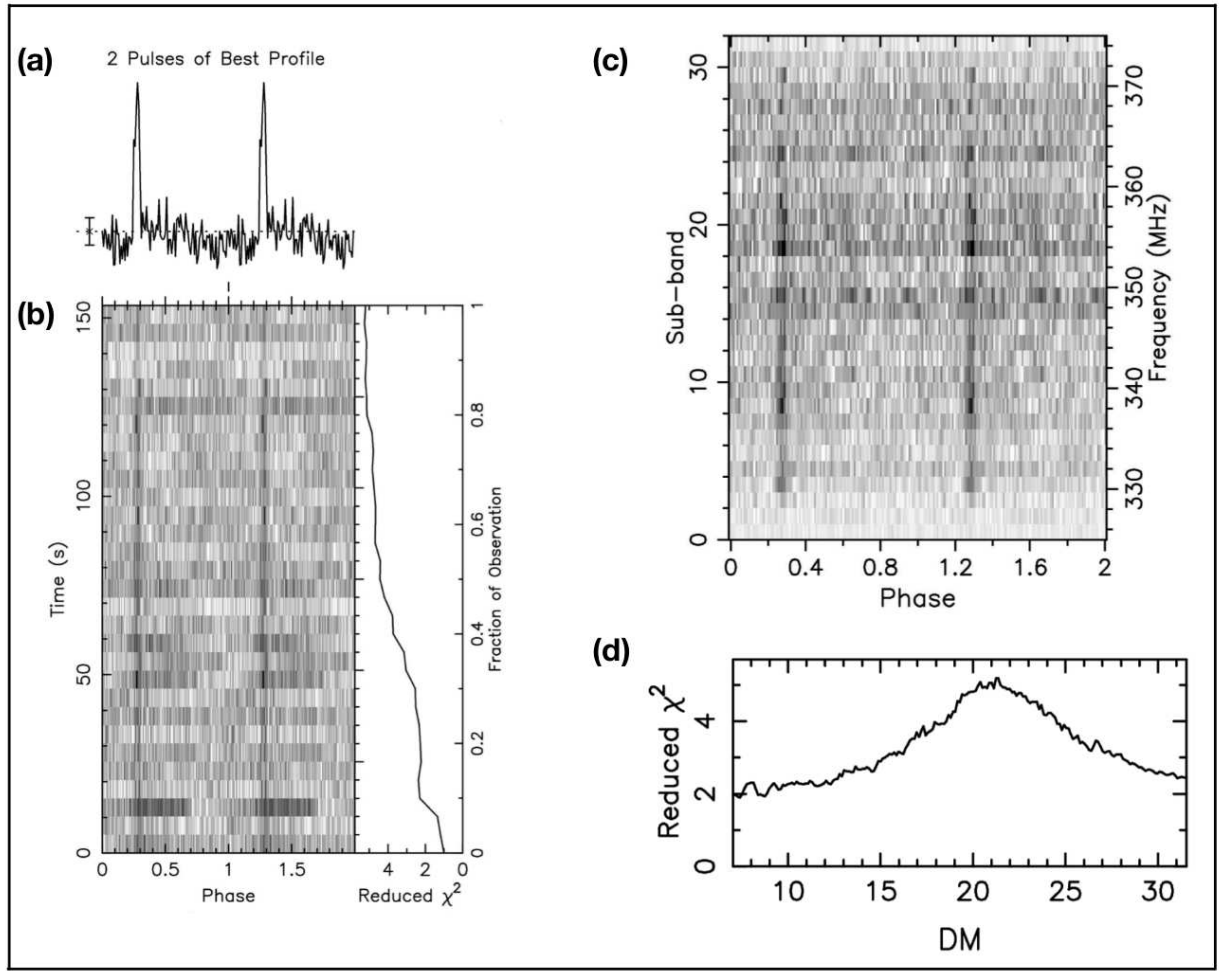}
\caption{An example of part of a periodicity search plot graded by PSC participants. (a) shows the pulse profile, (b) the time series, (c) the pulse profile vs. observing frequency/sub-band, and (d) the reduced $\chi^2$ vs. DM. Additional description is provided in the text.} \label{fig:fft}
\end{figure*}

After the data are de-dispersed, search algorithms are run to produce two different types of diagnostic plots: periodicity search plots and single-pulse plots. This paper will focus on periodicity search plots. In a periodicity search, for each trial DM, a Fast Fourier Transform (FFT) is applied to the time series to create a power spectrum. Peaks in the Fourier spectrum identify possible pulsar candidates. The time series is then ``folded'' (many pulses combined together to build up a detectable signal) at the period of each candidate to produce pulse profiles by summing segments of the pointing separated by the identified period. This sharpens the signal and reduces noise. 

Figure \ref{fig:fft} shows an example of part of a periodicity search plot. The full plot can be found in Williamson et al. (2019). A detailed explanation of periodicity search plots or single-pulse plots is provided at the PSC program website (\url{http://pulsarsearchcollaboratory.com/psc-manuals/}).  The different panels in Fig. \ref{fig:fft} present information that allows the identification of pulsars. The plot also provides an illustration of the pre-processing described above.  

Fig. \ref{fig:fft}(a) shows a pulse profile, which is the folded signal from a pulsar collected over the entire observation.  The signal is plotted twice for clarity and the sharp peaks signify a clear periodic signal. The $x$-axis in Figs. \ref{fig:fft}(a) to (c) is labeled as ``Phase'', which is the time taken by the pulsar to undergo one rotation. As such, Phase = 1 corresponds to one full rotation. The time-series plot in Fig.~\ref{fig:fft}(b) shows the strength of the signal as a function of phase and time (in seconds) since the start of observation. The greyscale level of each bin in a single row represents the strength of the radio signal; darker bins represent stronger signals. Hence the dark vertical lines in Fig.~\ref{fig:fft}(b) show a sudden increase in the strength of our signal (due to the pulsar beam passing in front of us) which occurred at the same phase. Fig.~\ref{fig:fft}(c) is similar to Fig.~\ref{fig:fft}(b), where the $y$-axis has been replaced with observing frequency (measured in MHz) or sub-band. The sub-band shows how our instruments are breaking down the wide range of frequencies into small chunks (here, 32 over the observation). The dark vertical lines in Fig.~\ref{fig:fft}(c) suggest that the signal is broadband, or detectable over all frequencies. Fig. \ref{fig:fft}(d) shows reduced chi-squared ($\chi^2$), or the significance, vs. DM, where the peak of the curve suggests the most likely value of DM.  A real pulsar has a peak value at non-zero DM, with the significance falling off smoothly on either side. 

\section{The PSC Program}
The PSC has been presented in two versions supported by separate NSF awards.  The first PSC award funded activities from 2008--2013 (hereafter, PSC I) and originally targeted students and teachers in and around West Virginia. During Fall 2015, WVU received another three-year NSF award. With this new grant, PSC II began in 2016 and the program expanded nationwide. Throughout the text, PSC refers to the program in general while PSC I and II specifically refer to the phases 2008--2013 and 2016--2019, respectively.

\label{sec:psc} This section introduces the PSC program. Some elements of the program have been consistent over its entire existence. These are discussed first in Section \ref{sec:genpsc}, followed by elements specific to the original regional version of the PSC (PSC I) and newer national version of the program (PSC II).

\subsection{The PSC}
\label{sec:genpsc}

Student and teachers in the PSC interact with two websites: the PSC program website and the PSC database website. The program website (\url{http://pulsarsearchcollaboratory.com}) contains a description of the program and training materials, and hosts a discussion forum that allows students, teachers, and astronomers to discuss pulsar science. The database website (\url{http://psrsearch.wvu.edu}) provides access to pulsar data; it contains diagnostic plots grouped into pointings that were produced by the data processing described in Section \ref{sec:data}. PSC students analyze pointings as a single package. It is also the site where students and teachers create their PSC accounts, where teachers manage their student accounts, and where students take their pulsar certification tests.

High school students use the graphical interface at the PSC database website to quantitatively grade pulsar plots. The PSC database provides different interfaces for different types of users such as teachers, students, and astronomers. Interested teachers join the program by registering and creating an online team which students then join. The student interface provides access to the diagnostic pulsar plots, which they examine to search for pulsars  after becoming eligible to review real data by passing two certification tests (see Section \ref{sec:psc2}). If a promising candidate is identified, they look the candidate up in a database of known pulsars. If the pulsar is not already known, they notify the astronomers that a possible pulsar has been found. A few times during the year, the PSC astronomers use additional observing time on the GBT (an event that is often presented online to the students) to take additional data to confirm that the candidate is a pulsar. In order to make the analysis more accurate, each dataset is reviewed five times by five different students. Once a pointing has been analyzed five times, the dataset is removed from the student interface. The teacher interface allows a teacher to deactivate students, choose a team leader, and review the pointings students have graded. An important feature of the PSC program is the encouragement for students working in teams. PSC students belong to a team and work on the project collectively rather than individually. In this way, group analysis and discussions are encouraged. The organization and structure of PSC teams sets the program apart from other online collaboration tools by having more personal and targeted interactions between the participants. 

\subsection{PSC I}

A detailed description of PSC I activities and outcomes was presented in Rosen et al. (2010); here, we provide a brief
overview of PSC I.  The components of PSC I included a Teacher Institute, Student Institute, and continuing academic year activities. During the Teacher Institute, teachers spent two weeks at the GBO to learn about the fundamentals of astronomy through a mini-course, attended research talks given by astronomers, and investigated simple research questions using a working 40-foot diameter radio telescope. During the Student Institute, students joined their teachers at the GBO for a six-day student orientation to PSC I. The Student Institute focused on specialized topics
about pulsars including observational techniques in radio astronomy, interpreting pulsar plots, and the use of the program and database websites. The teachers also practiced their newly acquired knowledge by leading most of these activities and lessons with their students. During the academic year, teachers introduced astronomical research and the PSC to their classes, described their experience as researchers, and recruited students to join school-based PSC clubs. After the club members passed the certification tests, they were given access to the PSC database.
The clubs met after school, during free time at school, or during class time to discuss and analyze data. Teachers and PSC students interacted with the astronomers through email, web-chats, webinars, forum postings to the PSC website, and by joining online observing sessions at the GBT. Each year PSC I culminated in a three-day Capstone seminar at WVU, where the PSC students shared their research through oral presentations, papers, or posters.

PSC I included 41 high schools with approximately 2500 students (roughly 60\% female and 50\% from rural areas) and 100 teachers. Students in PSC I discovered seven pulsars (including one double neutron star system and one millisecond pulsar) and detected 25 already known pulsars by analyzing 20\% of the data collected in 2007 that was set aside for the PSC program. Many students were authors on papers in scientific journals (Rosen et al. 2013; Swiggum et al. 2015). The reasons cited by students for their engagement in PSC were the ability to work with their friends in a fun learning environment, the intrinsic enjoyment that comes from thinking about astronomy and pulsars, the chance to visit others and present their work at the PSC's annual Capstone events on college campuses, praise and encouragement from their teacher, and the feeling of being part of a team that is really contributing to science (Rosen et al. 2010).

\begin{figure*}[th]
\includegraphics[width=\textwidth]{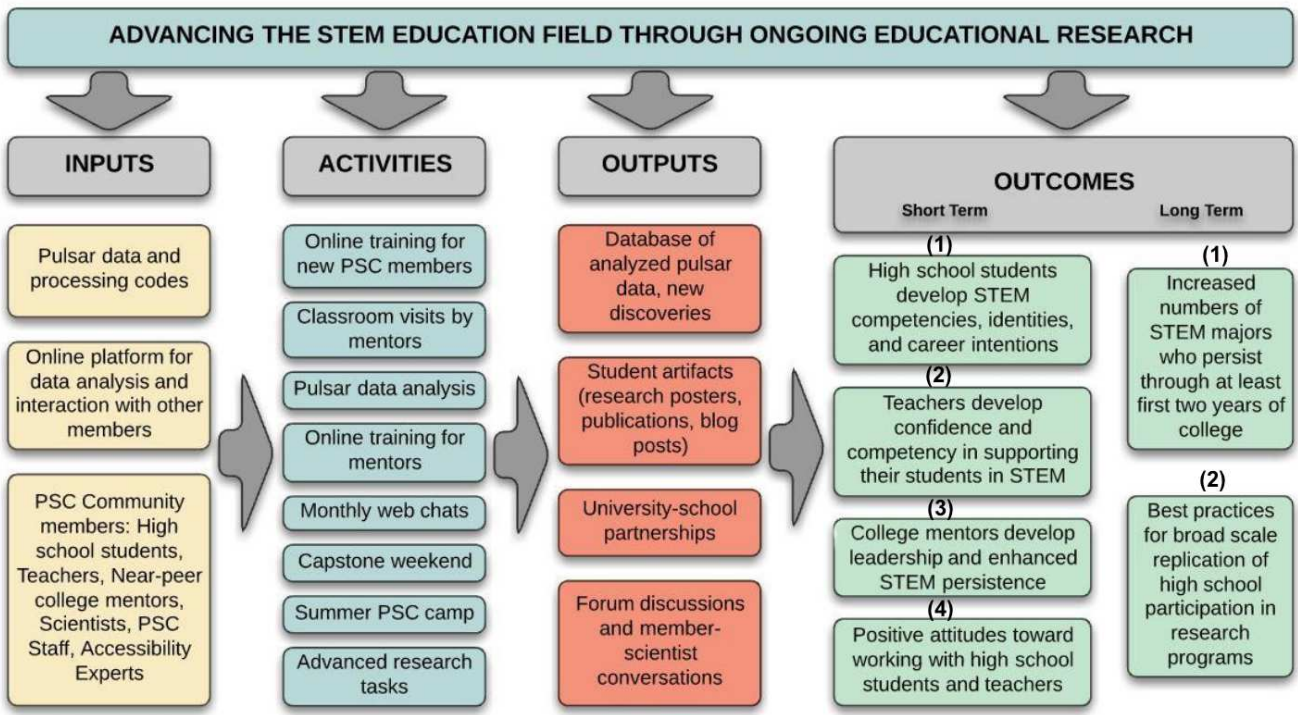}
\caption{Logic model for PSC II program showing the short-term and long-term anticipated outcomes, or goals. These are numbered for reference in the text. \label{fig:logic} }
\end{figure*}

\subsection{PSC II} \label{sec:psc2}

PSC II introduced a number of substantial programmatic innovations over PSC I: (i) a significant geographic expansion to partner institutions, called ``hub'' institutions, across the country, (ii) a more focused effort to include underrepresented groups, (iii) an online learning and discovery environment allowing the participation of students throughout the country, (iv) advanced research opportunities for deeper student participation in research and advancement within the research community, (v) mentoring by near-peer undergraduate students to build strong
bonds with high school students, and (vi) ongoing educational research informing programmatic modifications. Figure
\ref{fig:logic} shows the changes adopted while also retaining the core elements of PSC I (i.e., professional development to gain mastery of basic research techniques, formation of school-based clubs, pulsar data analysis, ongoing encouragement from experts, and sharing of results at Capstone Seminars located at a university). A more general overview of PSC II can be found in Williamson et al. (2019).

\subsubsection{PSC II data}

During the summer of 2016, the GBT surveyed the sky for an additional 375 hours using the 820-MHz receiver. In addition to detecting pulsars, this frequency is also sensitive to transient signals such as Fast Radio Bursts (FRBs), which are among the newest and most puzzling astronomical phenomena. FRBs are characterized by millisecond durations, large radio luminosities, and large dispersion measures (DMs); they are likely located at cosmological distances (Lorimer et al. 2007). The origins of FRBs are still unknown and their discovery has opened an entirely
new radio astronomy research area with high scientific potential. The study of FRBs will benefit enormously from the identification of additional sources.

\subsubsection{PSC II project design}

The PSC II program involves staff and scientists from WVU and GBO, faculty from hub institutions, undergraduate students, and high school students and teachers from across the nation. The undergraduate students act as mentors for the high school students. They reach out to regional and local schools, present talks on PSC and pulsar astrophysics, and invite high school students and teachers to participate. Interested students and teachers join the program. Some form PSC clubs at their schools while others work independently. The program is open to
any high school student (13 years and older) and teacher in the US. Students and teachers are recruited through an online application form, word-of-mouth, email flyers distributed to state science teacher listservs and NASA Educator Resource Center listservs, mentor school visits, outreach events, and presentations during conferences aimed at high school teachers.

The main components of PSC II include the following:

\begin{itemize}
\item{\it Online workshop}: All new members are required to participate in a six-week online training program offered each fall and spring on pulsar research and pulsar data analysis. The participants may participate in the training either synchronously by joining a live online training webinar or asynchronously using online video tutorials. Following
the online workshop, each new participant must pass certification tests where they demonstrate they can correctly identify pulsars and determine if a pulsar has already been discovered.
\item {\it Online environment}: The PSC program website provides access to training videos, interactive tutorials, animations, and data analysis materials. This site also hosts an online forum where all the PSC members are involved in proactively growing an online community by interacting with scientists, discussing research and data analysis
related topics, college life, and tips for success in STEM.   
\item {\it Annual Capstone seminar}: At the end of every academic year, some hub institutions host a Capstone seminar for students who have completed the online training and tests, and have graded 50 pointings of data. During this event, students and teachers present their research to other participants, attend talks given by scientists, network with each other, and experience campus life. 
\item {\it PSC Summer Camp}:  Each year during the summer, the PSC Camp is held at the GBO for 40 high school students, 20 teachers, and up to 16 undergraduate mentors. The camp is targeted at participants who have completed the online training, have completed the certification process, and have graded 50 pointings of data.  At the camp, students and teachers learn advanced topics in pulsar astronomy, pulsar searching, data reduction, and data analysis
techniques. They participate in leadership activities that prepare participants to take a larger role within the PSC and collaborate on the development of online tutorials for conducting more complex research tasks during the academic year. The students are split into small groups and given some short projects related to pulsars, which are to
be completed during the camp. At the end of the camp, all the groups present their results to the other participants.
\end{itemize}


\section{Results and Discussion}
\label{sec:results}

In order to assess the impact of PSC II and progress made toward reaching the stated outcomes in Fig. \ref{fig:logic}
pre- and post-surveys were administered to high school teachers, students, and undergraduate mentors. The WVU Institutional Review Board approved the research program and informed consent was collected from all participants. Different surveys were given to teachers, mentors, and students. For example, the high school teachers answered questions about their confidence in conducting research, skills in teaching science, and the effectiveness of
undergraduate mentors. The undergraduate mentors answered questions about their interest in STEM careers, confidence in their ability to conduct scientific research, their communication, leadership, and collaboration skills, their perception of their effectiveness as an undergraduate mentor, and the effectiveness of faculty mentors. The high school students were surveyed both at the end of the Capstone events and at the end of PSC camp. All respondents answered questions regarding the usefulness of program activities and their motivation to participate in the program.

PSC II participants at all levels were asked to complete survey questions scored on a 5-point Likert scale with larger responses representing more favorable responses. Differences in composite scores between reflective pre- and post-means were tested using $t$-tests. The practical size of pre/post differences was characterized using Cohen's $d$, the difference in the pre-survey and post-survey means divided by the pooled standard deviation. Cohen characterized $d>0.2$ as a small effect, $d>0.5$ as a medium effect, and $d>0.8$ as a large effect. Open response questions
were also administered. PSC II participants were 50\% male and 72\% White Non-Hispanic.

\subsection{High school teachers}

\begin{figure}

\includegraphics[width=0.45\textwidth]{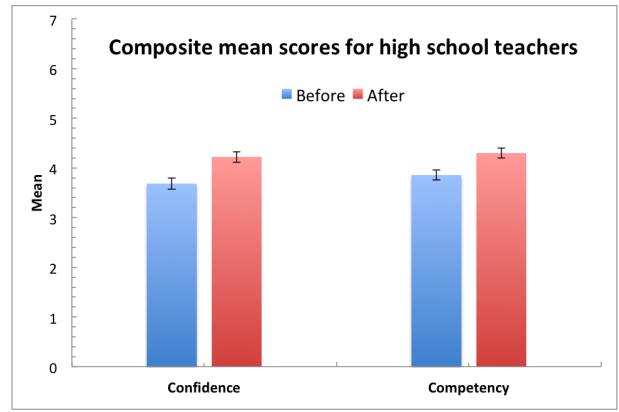}
\caption{Composite mean scores in confidence (left) and competency (right) for the high school teachers before
and after participating in the PSC program. All values measured on a 5-point Likert scale. \label{fig:teacher}}
\end{figure}

One of the main goals of the program was to deepen high school educators' self-efficacy and confidence in conducting scientific research and in their science teaching. Of the 141 high school teachers participating in PSC II, 64 completed the pre- and post-survey, a 45\% response rate. Teachers who participated for multiple years may have taken the surveys multiple times. These teachers were 51\% male and 95\% of the teachers identified as White non-Hispanic, mirroring the low racial and ethnic diversity of West Virginia. High school teachers rated their confidence in their ability to conduct research alongside scientists and their competence in teaching science before and after participating in
the PSC II program. Figure \ref{fig:teacher} shows the mean scores and standard errors.  After participating in PSC II activities, high school teachers showed a significant increase in their confidence in conducting research ($p<0.001$, $d=0.63$). This difference represents a medium effect size (Fig. \ref{fig:logic}, short term goal 2). Fifty-three percent of teachers in year 3 reported implementing astrophysics science/research into their classroom and 17\% indicated that they plan to implement it next year. Across all three years of the project, teachers consistently noted that they planned to implement PSC research in their classroom.

High school teachers also showed significant increases in their competency in teaching science and teaching the research process as shown in Fig. \ref{fig:teacher}, ($p=0.001$, $d=0.53$), also a medium effect (Fig. \ref{fig:logic}, short term goal 2).

\subsection{Undergraduate mentors}

\begin{figure}

\includegraphics[width=0.5\textwidth]{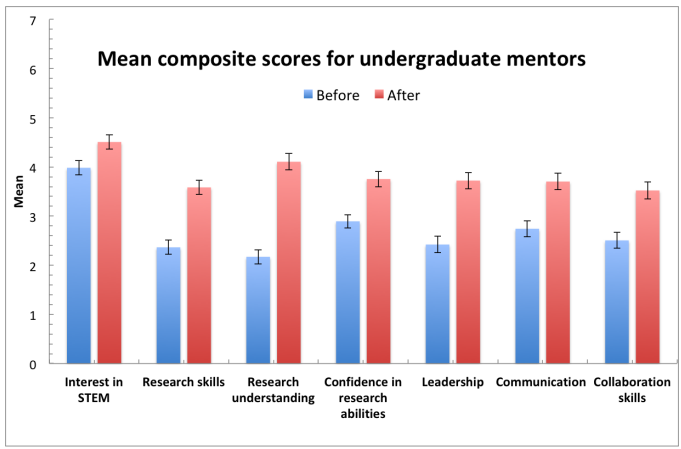}
\caption{Composite mean scores of the undergraduate student mentors before and after participating in the PSC II program. All values measured on a 5-point Likert scale. \label{fig:mentor}}
\end{figure}

For the undergraduate student mentors, the main program goal was to improve their persistence in STEM majors through the development of competencies in the 21st century skills. It should be noted that not all participants answered all questions on each survey and, therefore, the sample size varies at different time points. Mentors were given surveys which measured their scientific self-efficacy toward a number of activities: observing with a radio telescope, analyzing data, using astrophysics software programs, understanding of pulsar applications, and searching for pulsars through a series of survey questions. Figure \ref{fig:mentor} shows the composite mean scores along with the standard error for the undergraduate mentors before and after participating in PSC II. All survey items were on a 5-point Likert scale. Of the 122 undergraduate mentors, 64 completed the pre- and post-survey, a 52\% response rate. Mentors who participated for multiple years may have taken the surveys multiple times. The mentors were 50\% male and 81\% of the mentors identified as White non-Hispanic.

Some of the highlights from survey results include:
\begin{enumerate}
\item As shown in Fig. \ref{fig:mentor}, the student mentors already had strong interest and future plans to be in STEM careers and to graduate with a STEM major. In post-surveys each year, many mentors shared that participating in the PSC II program confirmed or increased their interest in astronomy and astrophysics, expanded their career options, and increased their interest in becoming a STEM educator and engaging in outreach activities.
\item The student mentors significantly increased their self-efficacy toward research skills ($p<0.001$, $d=1.5$) and
research understanding ($p<0.001$, $d=2.2$), very large effects. This indicates that PSC II was successfully impacting their scientific knowledge self-efficacy (Fig. \ref{fig:logic}, short term goal 3).
\item The student mentors indicated that their leadership and communication skills greatly improved ($p<0.001$,
$d=1.0$ and $d=1.3$, respectively), suggesting that PSC II was successfully impacting their communication and leadership skills (Fig. \ref{fig:logic}, short term goals 3 and 4).
\end{enumerate}

\subsection{PSC Capstone}

The high school student participants reflectively rated their agreement before and after the Capstone event with three
statements: (i) I believe I am part of a community of scientists, (ii) I believe I am advancing science through PSC research, and (iii) I have an interest in science-related careers. Of the 66 students who attended a capstone event, 59 completed the survey, an  89\% response rate. Overall, 83\%--97\% of the respondents agreed with all statements before participating in capstone. After Capstone, 93\% believed they were part of a community of scientists and 97\% agreed they are advancing science through PSC research. High school students shared that the most beneficial
things they learned were career/academic paths, understanding the importance of failure in research, interacting with other students and researchers, and learning about pulsar research (Fig. \ref{fig:logic}, short term goal 1). Undergraduate students shared that they enjoyed working with high school students. All participants further indicated that they would be able to implement the knowledge and skills they gained in their academic studies, career, and/or research.

\subsection{PSC Summer Camp} Students and teachers were also given a survey before and after participating in the
PSC summer camp. Of the 127 students and teachers who attended PSC camp, 113 completed the survey, an 89\% response rate. After participating in the summer camp, 85\% of the participants had increased understanding and competency in all areas, with the greatest increase in their ability to analyze plots and their understanding of radio astronomy (Fig. \ref{fig:logic}, short term goals 1--3). These findings suggest that the summer camp is successfully impacting participants' scientific knowledge in pulsar science and radio astronomy. High school student respondents had the greatest increase in their understanding of radio astronomy, whereas teacher respondents had the greatest increase in their ability to analyze plots. Eighty percent of the participants also felt they were part of a team following the PSC camp. In addition, several high school students and teachers shared that the most useful part of the camp was being able to learn about new career fields and being about to talk to graduate students and researchers during the activities.

\subsection{Sample quotes and responses}

While quantitative research is helpful, qualitative research is also needed to fully understand the PSC experience. Multiple interviews with participants of all levels (teachers, students, student mentors, and hub faculty) as well as two focus group activities yielded valuable insight into the program. We share below some sample questions and responses from the most recent focus group activity at the PSC summer camp.

\begin{enumerate}
\item ``What kept you motivated to continue participating in the PSC?'' --- The motivating aspects of PSC II appear to be much the same as those listed by students in PSC I, namely: the chance to make a discovery and participate in real science, interest in the astronomy and pulsars, and having fun learning with friends.
 \item
``What kind of support did you receive to prepare for PSC camp?'' --- Students reported high levels of self-directed engagement with the PSC and support from family or a particular teacher, i.e. most did not appear to belong to a club environment. Student quotes: ``My family and friends supported me but sadly my school didn't," ``Me, myself, and I?'',  and ``the [PSC] staff and the mentors, they were and have been patient.''
 \item ``Do you have advice for how to get others engaged in the PSC?'' --- Most feel like it is hard to adequately convey the excitement of PSC to others. All types of participants (hub leaders, teachers, staff, mentors,
students) agree that there should be more advertising. PSC student quotes: ``Advertise! I have many friends that love science and they would love to hear about this,'' ``Reach out to schools more, like high school and middle school, also more free food,'' ``Make it more known! It's a cool thing; people just don't know about it.''
\end{enumerate}

\subsection{Long term goals}

While it has been challenging to track PSC students participating across the country through college, the excitement toward radio astronomy generated by the program has resulted in a increase in the number of physics majors pursuing astronomy research at WVU (Fig. \ref{fig:logic}, long term goal 1). The experience and challenges of transitioning  a regional, largely face-to-face program to a national, largely online program have been instructive. The PSC program has shown the importance of the use of focus groups to fully understand the online experience, the importance of providing asynchronous online experiences for busy students and teachers, and benefits of careful user testing of online elements (Fig. \ref{fig:logic}, long term goal 2).

\section{Summary and future direction}

In summary, the opportunity to conduct pulsar research is a substantial benefit of PSC for all participants. High school
students and undergraduate mentors both felt that they had gained research skills and confidence.  Working with high school students and teachers was a strong motivating factor for undergraduate students. The experience helped improve their leadership and collaboration skills. Furthermore, participating in the PSC program helped high school teachers improve their competency and confidence in teaching science. The PSC Summer Camp continues to
be an important component of PSC and all respondents rated it the most useful component of the program.

The PSC II program is currently in its third year and includes approximately 500 high school students, 100 high school teachers, 80 undergraduate mentors, and 31 faculty mentors from 195 middle and high schools and 16 colleges and universities across the country.  This number is expected to increase over the next year. While PSC II students have not yet discovered a pulsar, several promising candidates have been identified; additional data will be taken to
determine if these candidates are real pulsars in the near future. A large amount of data ($\sim$80\%) still remains to be analyzed. Exciting discoveries from the 820-MHz survey data await. We expect higher scientific returns in the coming years, similar to the discoveries made by high school students in PSC I who were co-authors in the Rosen et al. 2013 and Swiggum et al. 2015 papers. The PSC continues to work with schools and teachers across the nation with the aim of positively impacting the students to support their STEM career ambitions. The PSC demonstrates that high school students can participate in authentic scientific research and that this can play a critical role in STEM interest and persistence. We encourage its use as a model for similar programs in other fields.

\begin{acknowledgments}

This work is supported by NSF Advancing Informal STEM Learning (AISL) (awards: 1516512, 1516269). The Green Bank Observatory is a facility of the National Science Foundation operated under a cooperative agreement by Associated Universities, Inc.

\end{acknowledgments}

\end{document}